 \documentclass[aps,twocolumn,prb,showpacs]{revtex4}

\begin{document}

\title{Optimal fluctuation approach to a directed polymer in a random medium}
\author{I.V. Kolokolov}
\affiliation{L.D. Landau Institute for Theoretical Physics, Kosygina 2,
Moscow 119334, Russia}
\author{S.E. Korshunov}
\affiliation{L.D. Landau Institute for Theoretical Physics, Kosygina 2,
Moscow 119334, Russia}
\date{\today}
\date{April 13, 2007}

\begin{abstract}

A modification of the optimal fluctuation approach is applied to study the
tails of the free energy distribution function $P_L(F)$ for an elastic
string in quenched disorder both in the regions of the universal behavior
of $P_L(F)$ and in the regions of large fluctuations, where the behavior
of $P_L(F)$ is non-universal. The difference between the two
regimes is shown to consist in whether it is necessary or not to take into
account the renormalization of parameters by the fluctuations of disorder
in the vicinity of the optimal fluctuation.

\end{abstract}

\maketitle

A large variety of physical systems can be described in terms of
an elastic string interacting with a quenched random potential.
The role of such a string can be played by a domain wall in a
two-dimensional magnet, a vortex line in a superconductor,
a dislocation in a crystal and so on,
however most often this class of systems is
discussed under the name of a directed polymer in a random medium.
The unfading interest to this problem is additionally supported by
its resemblance to more complex systems with quenched disorder
(e.g., spin glasses),
as well as by its relation to the dynamics of a
randomly stirred fluid and to stochastic growth (see Refs.
\onlinecite{Kardar-R} and \onlinecite{HHZ} for reviews).

A number of properties of a random directed polymer can be easily
deduced as soon as one knows $P_L(F)$, the free energy distribution
function for large polymer length $L$.
In 1987 Kardar  \cite{Kardar} proposed a  method for
calculating the moments of $P_L(F)$ in the one-dimensional (1D) problem
with $\delta$-correlated potential.
However, later it was understood  \cite{MK} 
that the approach of Ref. \onlinecite{Kardar} allows one to find
\cite{Zhang} only the tail of $P_L(F)$ at large negative $F$
(the left tail), so
the conclusions on the width
of this distribution function have to rely on
the assumption that at large $L$ it acquires a universal form,
 \begin{equation}                                     \label{P*}
 P_L(F)=P_*(F/F_*)/F_*\,,
 \end{equation}
incorporating the dependence on all parameters 
through a single characteristic free-energy scale $F_*(L)\propto
L^\omega$, which therefore can be extracted from the form of the tail.
Here and below it is assumed that $F$ is counted off from its average.

In the present work we report the results of the first systematic
investigation of $P_L(F)$ in the directed polymer problem which makes a
distinction between the interval of the universal behavior and the regions
of large fluctuations, where 
$P_L(F)$ can be expected to deviate from Eq. (\ref{P*}).
We demonstrate that the far-left tail of $P_L(F)$
is determined by the optimal fluctuation and apply an analogous approach
for the analysis of the far-right tail \cite{BFKL}.
Inside the universal region, we rely on the
generalization of this method accounting for the renormalization
of different parameters by fluctuations.

This allows us to show that in the 1D 
case the exponent $\eta_+$ in the dependence
\makebox{$\ln P_L(F)\propto -|F|^{\eta_+}$} is changed from 3
in the universal part of the right tail to $5/2$ outside of it
(at the largest $F$).
On the other hand, we find that in the left tail
the behavior of $P_L(F)$ is qualitatively the same
(with $\eta_-=3/2$) in both regions and
that the analysis of Ref. \onlinecite{Zhang} reproduces a
correct estimate for $F_*(L)$ only as a result of {this coincidence},
because in reality
it describes the {\em non-universal} part of the left tail.
We also analyze how the tails of $P_L(F)$ are modified when the
distribution of disorder is characterized by a finite correlation length.

{\em The model.} -
We consider an elastic string confined to a plane and interacting
with a random potential $V(t,x)$ with zero mean and Gaussian statistics
described by
$$
\overline{V(t,x)V(t',x')}=\delta(t-t')U(x-x')\,.
$$
An overbar denotes the average with respect to disorder and
$t$ is the coordinate along the string.
Such a string can be described by the Hamiltonian,
\begin{equation}                                          \label{H}
H 
=\int_{0}^{t}dt' \left\{
\frac{J}{2}\left[\frac{dx(t')}{dt'}\right]^2+V[t',x(t')]\right\} \;.
\end{equation}
The partition function of a string
which starts at $t=0$ and ends at the point $(t,x)$
is denoted $z(t,x)$ and  its free energy $f(t,x)=-T\ln z(t,x)$,
$T$ being the temperature.

Derivation of $f(t,x)$ with respect to $t$ shows \cite{HHF} that the
evolution of $f(t,x)$ with the increase in $t$ is governed by
the Kardar-Parisi-Zhang (KPZ) equation \cite{KPZ}
(with the inverted sign of the non-linear term),
\begin{equation}                                  \label{KPZ}
f_t+(1/2J)f_x^2-\nu f_{xx}=V(t,x)\,,
\end{equation}
where $t$ plays the role of time,
subscripts $t$ and $x$ denote partial derivatives and
$\nu={T}/{2J}$. 
On the other hand,
derivation of Eq. (\ref{KPZ}) with respect to $x$
leads \cite{HHF} to the Burgers equation \cite{Burgers}
with random force,
\begin{equation}                                 \label{Burg}
u_t+u u_{x}-\nu u_{xx}=V_x(t,x)/J\,,
\end{equation}
in which $u(t,x)=f_x(t,x)/J$ plays the role of velocity.

{\em The replica approach} to the directed polymer problem
is based on calculating the moments, $Z_n\equiv \overline{Z^n}$,
of the distribution of the partition function, $Z\equiv z(L,0)$,
and does not rely on the analytical continuation of $n$ to 0.
Kardar \cite{Kardar} has noticed that for any integer $n\geq 1$ (and
large enough polymer length
$L$) $Z_n$ with an exponential accuracy
can be approximated by $\exp[-E_0(n)L/T]$, where
$E_0(n)$ is the ground state energy of
$n$ bosons whose mass is equal to $J/T^2$ and interaction
to $-U(x)/T$. On the other hand,
since $Z^n\equiv\exp(-nF/T)$, ${Z_n}$ can also be expressed in
terms of $P_L(F)$, the distribution function of the free energy
$F\equiv f(L,0)=-T\ln Z$:
\begin{equation}                                       \label{expF}
Z_n=\int_{-\infty}^{+\infty}dF\,P_L(F)\,\exp(-nF/T)\,.
\end{equation}

In the case of $\delta$-correlated random
potential, \makebox{$U(x)=U_0\delta(x)$},
$E_0(n)$ can be found exactly \cite{Thacker}. This gives
\makebox{$Z_n\propto\exp[({JU_0^2}/{24T^5})n^3L]$}, where the linear in
$n$ term in $E_0(n)$ has been omitted, because it can be eliminated
by a constant shift of the potential $V(t,x)$ in Eq. (\ref{H}).
Comparison of the two expressions for $Z_n$
allows one to conclude \cite{Zhang} that for large negative fluctuations
$P_L(F)\propto \exp[-S(F)]$, where
\begin{equation}                                    \label{SF}
S(F)\approx\frac{2}{3}\left(\frac{-F}{F_*}\right)^{3/2},~~
F_*(L)=\frac{1}{2}\left(\frac{JU_0^2L}{T^2}\right)^{1/3}.
\end{equation}

For such $P_L(F)$ the integral in Eq. (\ref{expF}) at  positive
$n$ is dominated by the saddle point at $F\approx -F_c(L)n^2$, where
\begin{equation}                                      \label{condF}
F_c(L)= F_*^3(L)/T^2= JU_0^2L/8T^4\,.
\end{equation}
%
%
Since the explicit expression for $Z_n$ is known only at integer
\makebox{$n\geq 1$} and cannot be analytically continued to
\makebox{$0<n<1$},
\cite{Kardar-R,MK}
the region of the applicability
of Eq. (\ref{SF}) is restricted to \makebox{$-F\gg F_c(L)$.}
With increase in $L$ this region is shifted to larger and larger values
of the ratio 
\makebox{$|F|/F_*\gg F_c/F_*\propto L^{2/3}$},
which shows that the replica approach cannot provide any information
on $P_*[F/F_*(L)]$, {\em i.e.} on the universal form of $P_L(F)$
at $L\rightarrow\infty$.

{\em The optimal fluctuation approach.} -
If $V(t,x)$ would be independent of $t$, for large enough $L$
the free energy of a polymer in a given disorder realization
would be given by
$F\approx E_{0}L$, where $E_{0}$ is the ground state energy of the
single-particle quantum-mechanical Hamiltonian,
\makebox{$\hat{H}=-({T^2}/{2J})\partial_x^2+V(x).$}
When $V$ is fluctuating with $t$, one can nonetheless consider a
question about the form of the most probable ``stationary"
({\em i.e.,} uniform along $t$)
fluctuation of $V$ which leads to the given negative value of $E_0$.
According to the analysis of Ref. 
\onlinecite{HL}, this form is given by
\begin{equation}                                        \label{Vx}
    V(x)=-\Lambda\int_{-\infty}^{+\infty}
    dx'\,U(x-x')\Psi^2(x'),
\end{equation}
where $\Lambda$  is an arbitrary constant
and $\Psi(x)$ is the localized solution of the
non-linear Schr\"{o}dinger equation, \makebox{$E_0\Psi=\hat{H}\Psi$},
in which $V(x)$ has to be replaced by the right-hand side of Eq. (\ref{Vx}).

In the case of $\delta$-correlated random
potential, \makebox{$U(x)=U_0\delta(x)$},
such a solution (the soliton) has a form \cite{HL}
\makebox{$\Psi_\Delta(x)=T/[(\Lambda JU_0)^{1/2}\Delta\,\cosh(x/\Delta)],$}
where the parameter $\Delta(F)=T[L/2J(-F)]^{1/2}$ can be called
soliton width.
Substitution of $\Psi_\Delta(x)$ into Eq. (\ref{Vx})
then gives
\begin{equation}                                \label{Vx2}
V(x)=-({T^2/J}{\Delta^2})\cosh^{-2}(x/\Delta)\;.
\end{equation}
In such a potential there exists only one level with a negative
energy \cite{LL-QM},
from where the condition for the
applicability of the relation $F\approx E_0 L$ is $-F\gg T$.

On the other hand, substitution of Eq. (\ref{Vx2}) into
the functional,
\begin{equation}                               \label{SV}
S=\frac{1}{2U_0}\int_{0}^{L}dt\int_{-\infty}^{+\infty}dx\, V^2(t,x)\,,
\end{equation}
determining the probability of $V(t,x)$ for \makebox{$U(x)=U_0\delta(x)$}
leads to exactly the same expression for $S(F)$
as follows from the replica analysis, see Eq. (\ref{SF}).
This proves that for $\delta$-correlated random potential the far-left
tail of 
$P_L(F)$ is entirely determined by the optimal fluctuation of
$V(t,x)$.
However, in order to find the range of the applicability of such
an approach one has to consider the influence of the fluctuations
of disorder in the vicinity of the optimal fluctuation.
This can be done most transparently in terms of the Burgers equation
representation.

{\em The Burgers equation approach.} -
It follows from the definition of $u(t,x)= f_x(t,x)/J$ that
the stationary solution of the Burgers equation (\ref{Burg}) with
the pumping potential $V(x)$ of the form (\ref{Vx2}) is given by
\begin{equation}                                          \label{ueta}
u_\Delta(x)=-\frac{T}{J}\frac{d}{dx}\ln\Psi_\Delta(x)
         =u_\Delta\tanh(x/\Delta)\,,
\end{equation}
where $u_\Delta=T/J\Delta$.
Thus, in terms of the velocity $u(t,x)$ the stationary soliton
looks like an inverted standing shock wave. \cite{comm}
Such solitons have been discussed (in various contexts)
in a number of works by Fogedby. \cite{Fogedby}

The description in terms of the Burgers equation allows one to understand
rather easily how the shape of the optimal fluctuation
evolves with time when one takes into account the initial condition.
Below we adopt the free initial condition for the string, $z(0,x)=1$,
which in terms of interface dynamics corresponds to the standard initial
condition \cite{HHZ} for non-stationary growth,  $u(0,x)=0$.

If one assumes that for $t>0$ the potential $V(x)$ is of the form
(\ref{Vx2}), then the solution of (\ref{Burg}) has to be close to
(\ref{ueta}) for not too large $x$, but must be close to zero at large
enough distance from the region around $x=0$ where the pumping is
localized. It is clear that this can be realized via the presence of two
travelling shock waves, which does not require any additional pumping and
therefore does not change $S(F)$. The velocities of these waves are
determined by their amplitude, \makebox{$v=\pm u_\Delta/2$}, so for $t\gg
\Delta/u_\Delta$ they will be located at $|x|\approx u_\Delta t/2\gg
\Delta$,  {\em i.e.}, relatively far from the soliton core. Accordingly,
in this regime $F\equiv J\int_{-\infty}^0 dx \,u(x)$ has to be close to
$-(J/2)u_\Delta^2 t$, as it has been assumed above. It is possible to
check that the imposition of the fixed initial condition [$x(t=0)=0$] also
does not lead to the change of $S(F)$.

In terms of the Burgers equation parameters (the viscosity $\nu=T/2J$ and
the pumping force intensity \makebox{$D=U_0/2J^2$}),
Eq. (\ref{SF}) can be rewritten as
\begin{equation}                                        \label{S4}
    S(F)=\frac{4}{3}\frac{\nu}{D}
    \left[\frac{2(-F)^3}{J^3L}\right]^{1/2}\,,
\end{equation}
From the nature of the optimal fluctuation approach it is clear that
this expression can be expected to be valid only when the soliton is so
narrow,
$\Delta\ll x_0$ (where \makebox{$x_0\sim\nu^3/D\sim T^3/JU_0$}), that
one can neglect the renormalization of  parameters by the nonlinearity.
Note that the constraint $\Delta\ll x_0$ is equivalent to
the condition \makebox{$-F\gg F_c(L)$}
defining the region of the applicability of the replica approach results.

At $-F\ll F_c(L)$ ({\em i.e.}, $\Delta\gg x_0$) one has to take
into account the influence of the fluctuations in the vicinity of optimal
fluctuation. Since optimal fluctuation is quasi-stationary (see below),
this  (up to a numerical factor) can be done by replacing all parameters
in Eq. (\ref{S4}) by their effective values at the scales of the order of
$\Delta$ (and zero frequency). However, it is well known that the
amplitude of the non-linear term in Eq. (\ref{KPZ}) (and, therefore,
coefficient $J$) cannot be renormalized \cite{MHKZ}
due to the Galilean invariance. On the other hand, in the 1D case the ratio
$\mu=D/\nu$ has to remain unchanged \cite{KPZ} as a consequence of
the fluctuation-dissipation theorem 
obeyed by Eq. (\ref{Burg}). \cite{HHF}
The coefficient $\mu$ describes the spectral density of the equilibrium
equal-time fluctuations of $u$, which is known to be
exactly the same as in the absence of the non-linearity. \cite{HHF}

Since Eq. (\ref{S4}) includes only invariant coefficients
\makebox{$\mu=D/\nu$} and
$J$, it can be expected to remain applicable not only for $-F\gg F_c(L)$,
but also (with some numerical correction) in the region of the universal
behavior, $-F\lesssim F_c(L)$.
From the other side the region of the applicability of this estimate
is  restricted by the constraint \makebox{$L\gg (\Delta^3/\mu)^{1/2}$,}
which is required \cite{NT} for the attainment of equilibrium at
the length-scales of the order of $\Delta$ and is satisfied as soon as
$-F\gg F_*(L)$ [{\em i.e.}, $S(F)\gg 1$].
The fulfillment of this condition is necessary for the applicability of
the above argument (based on known properties \cite{KPZ,MHKZ} of an
{\em equilibrated} system) in our non-stationary situation.

{\em The right tail.} -
The very special shape [in terms of $f(t,x)$] of the optimal fluctuation
corresponding to the left tail originates from the possibility
to have a growing fluctuation inside of which almost everywhere
\makebox{$f_t+(1/2J)f_x^2\approx 0$,}
and therefore only a small part of the volume of this fluctuation
contributes to the functional (\ref{SV})
[where $V(t,x)$ should be replaced by the left-hand side of Eq. (\ref{KPZ})].
It is clear that in the case of the right tail
such a situation is impossible because $f_t$ has to have the same sign as $f_x^2$.
As a consequence, the optimal fluctuation corresponding to the right tail
must have a more simple shape which
can be characterized by a single relevant length-scale, $\Delta$.
This length-scale can be estimated from $f_t\sim (1/2J)f_x^2$, which gives
$\Delta(F)\sim (FL/J)^{1/2}$
and a temperature-independent estimate for $S(F)$,
\begin{equation}                                \label{RT-S2}
S(F)\sim \frac{F^{5/2}}{U_0J^{1/2}L^{1/2}}\;\,.
\end{equation}
Note that Eq. (\ref{RT-S2}) contains the unrenormalized coefficient $U_0$,
whereas the estimate for $\Delta(F)$ 
shows that it grows with increase in $F$.
This suggests that even for large $F$
it may be necessary to correct Eq. (\ref{RT-S2}) by taking into account
the renormalization of $U_0$.

It follows from the analysis of higher-order diagrams \cite{LLPP}
that in a stationary 1D system
the scale dependence of $\nu$ and $D\equiv U_0/2J^2$  can be found from the
self-consistent one-loop theory \cite{BKS,HF}(mode coupling approximation),
which for \makebox{$\Delta\gg x_0$} gives \cite{HF,NT}
\makebox{$\nu(\Delta)\sim(\mu\Delta)^{1/2}$},
\makebox{$D(\Delta)\sim(\mu^3\Delta)^{1/2}$}.
However, in the case of the
right tail the replacement of $D$ by $D(\Delta)$ (where $\Delta$ is the
optimal fluctuation width) overestimates the renormalization of $D$.  It
is so because the only source for the renormalization of $\nu$ and $D$ is
the non-linear term in Eq. (\ref{Burg}), and therefore this
renormalization can be effective only up to a length-scale $R\sim\mu JL/F$
at which the typical velocity of equilibrium fluctuations, $u_R\sim
(\mu/R)^{1/2}$, becomes comparable with the velocity $u_F\sim
F/J\Delta(F)$ characterizing the optimal fluctuation being considered.
For the left tail $R\sim\Delta$, whereas for the right tail
$R\ll\Delta$, so in the latter case the effective value of $D$ has to be
estimated as \makebox{$D(R)\sim\mu^2(JL/F)^{1/2}\ll D(\Delta)$}, from where
\begin{equation}                               \label{RT-S3}
    S(F)\sim \frac{F^{5/2}}{D(R)J^{5/2}L^{1/2}}\sim \left(\frac{F}{F_*}\right)^3\,.
\end{equation}
The conditions for the applicability of Eq. (\ref{RT-S3})
describing the universal part of the right  tail are
$S(F)\gg 1$ and $R\gg x_0$, which means $F_*(L)\ll F\ll F_c(L)$.
Note that here both $F_*(L)$ and $F_c(L)$ are
the same as in the left tail,
{\em i.e.} are given by Eqs. (\ref{SF}) and (\ref{condF}) respectively.
Another similarity to the left tail is that the fulfillment of the condition
$F\gg F_*(L)$ ensures that
\makebox{$L\gg \Delta^2/\nu(\Delta)\sim (\Delta^3/\mu)^{1/2}$,}
and therefore time $L$ is sufficient for attaining the equilibrium
at the length-scales of the order of $\Delta$. \cite{NT}

At $F\gg F_c(L)$ one gets $R\ll x_0$, which means that the renormalization
of $\nu$ and $D$ becomes completely ineffective. Therefore,
the far-right tail is described by Eq. (\ref{RT-S2}) with unrenormalized
parameters.

{\em A finite correlation length}. - If correlations of random potential
are characterized by a finite correlation length $\xi\equiv U_0/U(0)\ll x_0$
[where now $U_0\equiv\int dx\,U(x)$], the expression (\ref{SF})
remains applicable only for $\Delta\gg\xi$, that is for
$-F\ll F_\xi\sim(x_0/\xi)^2F_c$.
In the other regime (at \makebox{$-F\gg F_\xi$}) the characteristic size
of $\Psi(x)$ becomes much smaller then $\xi$, which leads \cite{ShEf}
to $V(x)\propto -U(x)$, $E_0\propto -V(0)$
and $S\approx F^2/2U(0)L$.  This means that the finiteness of
$\xi$ makes the most distant part of the non-equilibrium left tail
Gaussian \cite{Gt} and independent of $T$.

With decrease of $x_0$ ({\em i.e.}, of temperature) the region with Gaussian
behavior 
becomes larger and larger, and for $x_0\ll\xi$
[{\it i.e.}, $T\ll T_0\sim(JU_0\xi)^{1/3}$] the whole non-universal tail
must be Gaussian. In this regime the parameters of
the universal part of the left tail also are modified.
For  $x_0\ll\xi$ the spectral density of the equilibrium fluctuations of
$u(x)$ at scales exceeding $\xi$ can be estimated as $u_\xi^2\xi$, where
\makebox
{$u_\xi\sim (D\tau_\xi/\xi^3)^{1/2}$}
is the characteristic velocity which is
created by random force with characteristic length-scale $\xi$
during the time $\tau_\xi\sim \xi/u_\xi$
required for the breaking of such a fluctuation.
This gives one an estimate \makebox{$\mu\sim (D^2/\xi)^{1/3}$},
substitution of which into Eq. (\ref{S4}) instead of the ratio $D/\nu$
leads to $S(F)\sim [F/\tilde F_{*}]^{3/2}$, where the 
free energy scale \makebox{$\tilde
F_{*}(L)\sim(x_0/\xi)^{2/9}F_*(L)\sim(JU_0^4/\xi^2)^{1/9}L^{1/3}$}
(replacing $F_*$) does not depend on $T$.
Directly in terms of the polymer problem an analogous estimate for
$\tilde F_*(L)$ can be obtained from scaling arguments complemented
by the assumption that at low enough temperatures $\tilde F_*(L)$ {\em has}
to be temperature independent, \cite{NR} and follows also from
the replica-symmetry breaking analysis of Ref. \onlinecite{KD}.

It can be shown that $\tilde F_*(L)$ replaces $F_*(L)$ also in
Eq. (\ref{RT-S3}) describing the universal part of the right tail.
On the other hand, 
the finiteness of $\xi$ does not change the form
of the non-universal part of the right tail described by Eq. (\ref{RT-S2}).
All this allows one to conclude that at
low temperatures $P_L(F)$ becomes temperature independent,
whereas the crossover between the regions of universal and
non-universal behavior takes place at
\makebox{$|F|\sim \tilde{F}_{c} 
\sim(U_0^2/J\xi^4)^{1/3}L$}.

{\em Conclusion. -} In the present work we have studied the form of the
free energy distribution function $P_L(F)$ in the 1D
directed polymer problem both for $\delta$-correlated random potential and
for a finite correlation length $\xi$. In terms of the KPZ problem
the same distribution function 
describes the distribution of heights in the process of
non-stationary growth starting from  flat interface.

Our analysis has shown that for $|F|\ll F_c(L)\propto L$
both tails of $P_L(F)$ are characterized by the same
free energy scale, $F_*(L)$, 
which confirms the universality hypothesis (\ref{P*}).
For  $T\gg T_0\sim(JU_0\xi)^{1/3}$ this scale is temperature dependent
[see Eq. (\ref{SF})], whereas at $T\ll T_0$ it saturates at a finite
value. 
On the other hand,
the exponents $\eta_-=3/2$ and $\eta_+=3$
in the dependence $\ln P_L(\pm F)\propto\! -(F/F_*)^{\eta_\pm}$
are insensitive to the relation between $T$ and $T_0$.
The same values of $\eta_-$ and $\eta_+$  have been found by
Pr\"{a}hofer and Spohn \cite{PSp} for  the polynuclear growth (PNG) model,
which in terms of a directed polymer problem, Eq. (\ref{H}), corresponds
to the Poisson distribution of idential point-like impurities
and a rather perculiar limit of $J=0$ and
$T=0$. \cite{PSp,Johanss}
This agreement confirms that the two models indeed belong to the same
universality class.

We also have investigated the non-universal tails of $P_L(F)$  (at $|F|\gg
F_c$). In particular, we have demonstrated that in the far-right tail
$\eta_+=5/2$, which is in agreement with the numerical results of Kim {\em
et al.}, \cite{KMB} who found $\eta_+=2.4\pm 0.2$. On the other hand, the
form of the far-left tail depends on $\xi$. For $\xi=0$ it is the same
($\eta_-=3/2$) as in the universal region, whereas for $\xi>0$ the left
tail becomes Gaussian for sufficiently large $|F|$.

The authors are grateful to V.B. Geshkenbein, A.I. Larkin and V.V. Lebedev
for numerous useful discussions.
The work of I.V.K. was supported by RFBR under grant 05-02-17305.



\begin{thebibliography}{99}
\bibitem{BFKL}    An analogous approach has been used \cite{GM} to study
                  the distribution of of velocity and its derivatives
                  in the Burgers turbulence problem,
                  which however requires to consider optimal fluctuations
                  with completely different structures then here.
\bibitem{comm}    This means that $u(x)=-u_\Delta(x)$ describes the genuine
                  standing shock wave which for any $\Delta$ is the solution
                  of Eq. (\ref{Burg}) with the vanishing right-hand side.
\bibitem{Gt}      This can be also shown in terms of the replica approach.
                  The tail becomes Gaussian when the size of
                  the ground-state wave-function of $n$ bosons becomes
                  much smaller than their interaction range $\xi$.
\end{thebibliography}
\end{document}